# Multiplexed multiple-τ auto- and cross- correlators on a single FPGA

Gábor Mocsár,[1*] Balázs Kreith,[1*] Jan Buchholz,[2] Jan Wolfgang Krieger,[2] Jörg Langowski[2] and György Vámosi[1§]

[1)]*University of Debrecen, Medical and Health Science Center, Faculty of Medicine, Department of Biophysics and Cell Biology, H-4032 Debrecen, Nagyerdei krt. 98, Hungary*
[2)]*German Cancer Research Center (DKFZ), Biophysics of Macromolecules (B040), Im Neuenheimer Feld 580,D-69120 Heidelberg, Germany*
[*]Equal contribution
[§]Correspondence should be addressed to G.V. (vamosig@med.unideb.hu)

**Abstract**

Fluorescence correlation and cross-correlation spectroscopy (FCS, FCCS) are widely used techniques to study the diffusion properties and interactions of fluorescent molecules. Autocorrelation (ACFs) and cross-correlation functions (CCFs) are typically acquired with fast hardware correlators. Here we introduce a new multiple-τ hardware correlator design for computing ACFs and CCFs in real time. A scheduling algorithm minimizes the use of hardware resources by calculating the different segments of the correlation function on a single correlator block. The program was written in LabVIEW, enabling computation of two multiple-τ ACFs and two CCFs on a National Instruments FPGA card (NI 7833R) in real time with a minimal sampling time of 400 ns. Raw data are also stored with a time resolution of 50 ns for later analysis. The design can be adapted to other FPGA cards with only minor changes and extended to evaluate more inputs and correlation functions.

## I. Introduction

### A. Fluorescence correlation spectroscopy (FCS)

FCS is widely used for studying diffusion properties and interactions of fluorescent molecules[1]. In this method a small volume element (~0.1 μm$^3$) is illuminated with a focused laser beam, and the time trace, $I(t)$, of the fluorescence intensity fluctuating due to molecules diffusing across the observation volume is recorded by photodetectors (typically avalanche photodiodes, APDs). The stochastic intensity fluctuations are characterized by the normalized autocorrelation function (ACF) of $I(t)$:

$$g(\tau) = \frac{\langle I(t) \cdot I(t+\tau) \rangle}{\langle I(t) \rangle^2} \qquad (1)$$

Angled brackets indicate time averaging over the whole measurement. Analyzing the ACF allows us to determine molecular properties, e.g. the diffusion coefficient of the fluorescent particle, the mechanism of diffusion, kinetics of molecular interactions affecting the fluorescence quantum yield of the dye, etc.[2-4]. FCS can detect significant changes of the size of the diffusing particle such as binding of a small molecule to a large structure. However, the dependence of the diffusion coefficient on the molecular mass is weak ($D \propto M^{-1/3}$), thus autocorrelation is relatively insensitive to small changes of the mass such as that occurring upon the association of molecules with similar size. For the detection of such events fluorescence cross-correlation spectroscopy (FCCS) is more appropriate. For FCCS, molecules are labeled with different colors. The fluorescence intensity from the observation volume is detected in two separate spectral intervals by two detectors. This allows calculating their cross-correlation function (CCF), which then contains information about those molecules carrying two different dyes. Analogous to Eq. 1, the normalized CCF (e.g. for the *x* and *y* detection channels) is defined as:

$$g^{xy}(\tau) = \frac{\langle I^x(t) \cdot I^y(t+\tau) \rangle}{\langle I^x(t) \rangle \langle I^y(t) \rangle}. \qquad (2)$$

Knowledge of the autocorrelation function of both channels and their cross-correlation functions can yield more detailed information about the molecular species, e.g. the fraction of molecules in complex relative to the total amount of molecules present[5-8].

Correlation functions are typically acquired in real time by dedicated hardware[9]. Field Programmable Gate Arrays (FPGAs) are reconfigurable hardware devices capable of performing parallel calculations such as the calculation of correlation functions. An approach to use an FPGA for calculating correlation functions of dynamic light scattering data is outlined in Ref.10 and 11. Our paper describes a new correlator design in detail, implemented on a National Instruments FPGA board using LabVIEW 8.6. It computes two ACFs and two CCFs simultaneously in real-time from photon streams detected by two APDs of an FCS microscope. Scheduler algorithms are described for calculating different segments of the correlation function(s) efficiently on a quasi-logarithmic time base (multiple-τ scheme). Minimal lag-times are 100 ns for calculating one ACF and 400 ns for calculating 2 ACFs and 2 CCFs. The design can be applied to calculate more than four correlation functions at the cost of increasing the minimal lag-time.

## B.  Hardware correlator

The purpose of a hardware correlator is to calculate the discretized form $\hat{g}(\tau_k)$ of the correlation function in real-time. Defined as:

$$\hat{g}(\tau_k) = \hat{g}(k \cdot \tau_{\min}) = \hat{g}_k = \frac{\overbrace{\frac{1}{T-k}\sum_{i=k}^{T-1} I_i \cdot I_{i-k}}^{G_k}}{\left(\underbrace{\frac{1}{T}\sum_{i=0}^{T-1} I_i}_{M}\right)^2}, \quad k = 1, 2 \ldots, T-1 \qquad (3)$$

where the photon counts $I_i$ are integrated over sampling periods of length $\tau_{\min}$ and the correlation function is determined over a set of lag-times $\tau_k$. The denominator describes the so-called *asymmetric normalization*[12].

FIG. 1 shows an implementation of a linear hardware correlator. There the signal $I^{(u)}$ (undelayed intensity) contains the current intensity $I_i$, and the signals $I_k^{(d)}$ represent different delayed input signals $I_{i-k}$. The different $G_k$ can be calculated using a multiplier and an accumulator (together called a correlation channel circuit). The delayed signals $I_k^{(d)}$ are organized as a shift register. The value of $M$ is calculated in the so-called monitor channel, which accumulates the undelayed intensities.

In typical FCS measurements the correlation function has to be computed over several decades of lag times, thus a linear spacing of $\tau_k$ ($\tau_k = k \cdot \tau_{\min}$) in a lag-time range of e.g. 1 µs…1 s and with $\tau_{\min} = 1$ µs would require $10^6$ correlation channels.

A more practicable choice is the multiple-τ scheme[12]. Here $S$ chained linear correlator blocks of only a few correlation channels each are combined. Each correlator block ($s=0,..,S$-1) contains $L$ delayed intensity channels $I_{s,0}^{(d)} + \ldots + I_{s,L-1}^{(d)}$ and $L$ correlation channels $G_{s,0}, ..., G_{s,L-1}$. After $m$ sampling periods of the $s$-th correlator block the last $m$ intensities ($I^{(d)}_{s,L+m-1} + ... + I^{(d)}_{s,L-1}$) are accumulated and propagated to the ($s$+1)-th correlator block as its new delayed intensity element ($I^{(d)}_{s+1,0}$). The undelayed intensity $I^{(u)}_s$ is also accumulated over $m$ cycles in block $s$ and the summed value is used to update $I^{(u)}_{s+1}$. The undelayed intensity $I^{(u)}_0$ of the first correlator block is updated with the photon count detected by the APD in the last sampling period. The delayed intensity channels inside block $s$ are updated by shifting the previous value of $I^{(d)}_{s,l}$ to $I^{(d)}_{s,l+1}$. Inside a correlator block the spacing of lag-times is linear; between consecutive blocks it increases by a factor of $m$. Within the $s$-th linear correlator block (referred to as the $s$-th block) the lag-times of the correlation channels are:

$$\tau_{s,l} = \begin{cases} l \cdot \tau_{min} & \text{if } s = 0 \\ \tau_{s-1,L-1} + \tau_{min} \cdot m^{s-1} \cdot (1 + l \cdot m) & \text{if } s > 0 \end{cases} \quad (4)$$

where $s$ is the index of the correlator block ($s = 0,1,...,S-1$) and $l=0,1,...,L-1$ is the index of a correlation channel within the block. Altogether there are $S \times L$ correlation channels.

The typical value of $m$ is 2, doubling the integration time from one block to the next, and each block is executed only after the previous block has been executed twice. Accordingly, the $s$-th block is executed only half as often as the preceding block.
In total, if the $s$-th block is executed $n_s$ times during the calculation, the overall number of executions after $k \cdot 2^{S-1}$ $(k \in \mathbb{N})$ periods will be

$$n_{total} = n_0 + n_1 + ... + n_{S-1} = n_0 \cdot (1 + \frac{1}{2^1} + \frac{1}{2^2} + ... + \frac{1}{2^{S-1}}) < 2 \cdot n_0 \quad (5)$$

Therefore, if the data input rate of the zero-th block is halved (i.e. it is executed every other cycle) and proper sequential scheduling of the execution of all other blocks is used, implementation of a single correlator block suffices to calculate all segments of the multiple-τ correlation function (see FIG. 2). In the next chapter we show the implementation of this design.

## II. Optimized multiple-τ correlator design

Our implementation of the multiple-τ scheme uses a single linear hardware correlator (referred to as correlator unit) to calculate all $S$ blocks. The random access memory (RAM) blocks of the FPGA are used to store the current state of the $s$-th correlator block ($G_{s,0...L-1}, I^{(d)}_{s,0...L-1+m}, I^{(u)}_s, I^{(u)*}_s, M_s, T_s$, definition of the latter quantities see below). A scheduler decides which correlator block should be calculated in a given cycle.

### A. Correlator Unit

The correlator unit performs the calculation of a selected linear correlator block, i.e., the $s$-th segment of the correlation function. An overview of the whole logic circuit is shown in FIG.3. There are $m$=2 additional delayed intensity channels $I^{(d)}_{s,L...L+m-1}$ and another undelayed channel $I^{(u)*}_s$. These channels are used for data handover to the next block. The state of the $s$-th correlator block is updated by the following operations:

1. The *undelayed intensity channels* $I_s^{(u)}$ contain the photon counts having arrived in the last sampling periods. The undelayed intensity $I_0^{(u)}$ for the 0-th block is received from an accumulator that is summing up the photon counts detected by the APD between two executions of this block. An additional undelayed intensity $I_s^{(u)*}$ stores the previous value of $I_s^{(u)}$ until the next sampling period. The undelayed intensity of a higher block is calculated as $I_{s+1}^{(u)} = I_s^{(u)} + I_s^{(u)*}$.
2. The values of the *delayed intensities* $I_{s,l}^{(d)}$ are shifted to the next channels $I_{s,l+1}^{(d)}$, $l \in [0, L+m-1]$. The last $m$ (2) intensities are so-called interleaved channels; the sum of the interleaved channels $I_{s,L}^{(d)} + I_{s,L+1}^{(d)}$ is passed to the $(s+1)$-th correlator block after every $m$-th execution of the $s$-th block, which defines the next value of the first intensity of the $(s+1)$-th correlator block $I_{s+1,0}^{(d)}$.
3. The values of $G_{s,l}$ are updated by adding $I_s^{(u)} \cdot I_{s,l}^{(d)}$, $l \in [0, L-1]$
4. The value of $T_s$ (number of executions of block $s$) is incremented by 1
5. The value of the monitor channel $M_s$ (initial value: 0) is updated by adding $I_s^{(u)}$.

Each time the correlator unit is executed with a new set of inputs, it updates all the $L$ correlation channels of the actual block (in our design $L=8$). A scheduler decides which correlator block is updated at a given time point.

**B.    Scheduler**

The scheduler ensures that blocks are executed in the right order, when input data from lower blocks are available. This means that the $(s+1)$-th block can be executed only after the $s$-th block has been executed twice since the previous execution of block $s+1$. For $m=2$ the sequence determining the index $s_c$ of the block to be executed in the $c$-th execution of the correlator unit can be written as

$$s_c = \begin{cases} 0 & \text{if } c \text{ is odd or } 0 \\ s_{c/2} + 1 & \text{otherwise} \end{cases} \quad c = 0,1,2,... \tag{6}$$

To avoid the recursion in Eq. 6, the following formula is used to define the same sequence:

$$\begin{aligned} s_c &= \log_2(X_c) \\ \text{with } X_c &= \left((c \text{ XOR } (c-1))+1\right)/2 \quad \text{if } c > 0, \text{ and } s_0 = 0 \end{aligned} \tag{7}$$

where XOR is a bitwise exclusive OR. Since $X_c$ is a power of 2, $s_c$ is always an integer.
In addition it is also checked if the shift register of the delayed channels of block $s_c$-1 is completely filled with data, which occurs after $L+m$ execution cycles of the block. If this condition is not fulfilled or if $s_c \geq S$, then no correlation is calculated in that cycle. The following equation gives the smallest value of $c$ for which the $s$-th block can be executed:

$$c_s^{(\text{threshold})} = 19 \cdot 2^s - 16 \tag{8}$$

(Proofs related to equations 6, 7 and 8 are in the Appendix A, B and C).

The combination of the correlator unit executing different linear correlator blocks sequentially and the scheduler calculating $s_c$ is called correlation processing element (CorrPE).

## C. Correlation Processing Element (CorrPE)

The CorrPE that calculates the multiple-τ estimate of the autocorrelation of one input signal is referred to as CorrPE-1. With appropriate modifications the CorrPE-1 can be extended to CorrPE-2X that estimates two ACFs and two CCFs. An overview of CorrPE-1 and CorrPE-2X is shown in FIG. 4a and b.

### 1. CorrPE-1

Each execution of a CorrPE-1 consists of the following operations controlled by an execution cycle counter $c$:

1. The scheduler calculates $s_c$ by Eq. 7.
2. The state of the $s_c$-th block is read from the RAM to the correlator unit.
3. The correlator unit updates the state of the block.
4. The new state of the block is stored in RAM.
5. $c$ is incremented by 1.

If the conditions for the execution of the correlation calculation of block $s_c$ are not fulfilled, (i.e., the required data are not available yet - see the section on the scheduler), then an empty execution occurs and only step 5 is carried out.

TABLE I shows the order of executions of the CorrPE-1 design.

### 2. CorrPE-2X

Instead of implementing four correlation units, one dedicated to each correlation function, our design calculates all correlation functions by a single correlator unit. Two accumulators are used for $I_0^{(u,x)}$ and $I_0^{(u,y)}$ (FIG. 4b). The states of two ACF and two CCF blocks with the same $s$ value are updated in four consecutive execution cycles for each combination of $x$ and $y$ (*xx, yy, xy, yx*) in a way analogous to that described for CorrPE-1. The states of all correlator blocks are also stored in the RAM of the FPGA.
To identify which correlation block (*xx, yy, xy* or *yx*) has to be calculated by the correlator unit, a periodic counter ($c_{corr}$) is used in a range of [0,1,2,3]. ACFs are executed for $c_{corr}$=[0,1] and CCFs for $c_{corr}$=[2,3]. Each execution cycle carries out the same operations 1-4 as those described previously for CorrPE-1; in step 5, $c_{corr}$ is incremented by 1. The value $c$ of the execution cycle counter is incremented only if all four correlation calculations within the same lag-time segment have been performed. TABLE II shows the order of execution of blocks with different values of $c_{corr}$ and $c$.

## D. Read-out and normalization

In both designs, all states for each correlator block are written to a dedicated address of the FPGA-RAM. After each execution of the last block the states of the correlation and monitor channels are copied to an intermediate location in FPGA-RAM and then to the DMA-FIFO (an element on the NI FPGA board responsible for direct memory access). The original memory locations storing the $G_{s,l}$, $M_s$ and $T_s$ values are reset to zero to prevent overflow. All parameters are read out from the DMA-FIFO by the host computer regularly with a period of $\tau_{\min} \cdot 2^s$, and accumulated in the host computer's memory. An additional DMA-FIFO is used to transmit the raw photon streams. We use asymmetric normalization (see Eqs. 3 and 9), which is done on the host computer. In order to estimate two normalized ACFs and CCFs of two input signals simultaneously (e.g. green and red channels) using the multiple-τ scheme, a discretized form of eq. 2 is used (for any *x-y* pair):

$$\hat{g}_{\text{asym,multi-}\tau}^{(x,y)}\left(\tau_{s,l}\right) = \frac{\frac{1}{T_s - l} G_{s,l}^{(x,y)}}{\frac{1}{T_s - l} M_s^{(x)} \cdot \frac{1}{T_s - l} M_s^{(y)}} = \frac{(T_s - l) \cdot G_{s,l}^{(x,y)}}{M_s^{(x)} \cdot M_s^{(y)}} \quad (9)$$

where *x* and *y* denote detectors. (*xx* and *yy* denote auto-, *xy* and *yx* cross-correlations). Division of $G_{s,l}^{(x,y)}$ and $M_s^{(x)}, M_s^{(y)}$ values by $T_s - l$ is required because out of the $T_s$ executions of block *s*, the first *l* executions yield zero for $G_{s,l}^{(x,y)}$ (the shift registers are not filled completely with photon counts). This process allows us to calculate and display normalized correlation functions on-line.

## III. Implementation

We implemented these designs on a Virtex-II (XC2V3000, Xilinx, http://www.xilinx.com/) FPGA residing on a PXI-7833R card (National Instruments), which was mounted in a PXI-1033 rack connected to the host PC via a PCI Express card.

The hardware correlator was developed in LabVIEW 8.6 (National Instruments) with the FPGA module. Both CorrPE designs operate at 20 MHz, thus the minimal sampling period is 50 ns. Since block 0 is executed on every other clock cycle, the minimal lag time is 100 ns in CorrPE-1 and 400 ns in CorrPE-2X. With this setup we can reach a maximal lag-time of 24 s using *S* = 25 correlator blocks for the CorrPE-1 and *S* = 23 for CorrPE-2X.
In both CorrPEs the parameters ($I_{s,l}^{(d)}, I_s^{(u)}, I_s^{(u)*}, G_{s,l}, M_s, T_s$) used by the linear correlators are stored in 31 RAM blocks (see TABLE III). Values of $G_{s,l}$ are stored as 64-bit words. All other parameters are 32 bit wide. For the read-out process 9 additional 64-bit RAM blocks are used. Additional RAM blocks are used by the DMA engine. The whole design reserves 60 FPGA RAM blocks out of 96 (62%). The resource consumption of the designs is shown in TABLE III and IV.

LabVIEW generates a *bitfile* which contains the FPGA configuration in binary form and can be directly loaded. Using the libraries provided by NI together with the bitfiles, an application uploads and runs the different compiled designs (CorrPE-1 and CorrPE-2X) on the FPGA card, and writes the data received from the DMA-FIFO into files. This application is synchronized with *ACCF Controller*, a Microsoft Visual C++ application, which normalizes and displays the correlation functions on-line.

## IV. Design validation and application to FCS

Our designs were tested with synthetic input data, showing correctness at the bit level. For testing with real data, FCS/FCCS measurements were carried out on an Olympus FV1000 confocal microscope equipped with a custom-designed FCS spectroscope employing 3 APDs (SPCM-APR-14, EG&G).
To test the ACFs generated by CorrPE-1, 50 nM solutions of Alexa Fluor 488 or Cy5 dyes were used as standards. For testing the CorrPE-2X a solution of W6/32 antibodies doubly labeled with Alexa Fluor 488 and Cy5 was used as a positive cross-correlation standard. As a negative control for cross-correlation a mixture of Alexa Fluor 488 and Cy5 dyes was applied.

Dyes were excited at 488 and 633 nm. Emitted photons were separated by a 570DXCR dichroic mirror, and filtered with 514/20 and 690/70 emission filters to detect them with $APD_x$ and $APD_y$, respectively. The TTL signals from the APD were split and fed simultaneously into an ALV5000E hardware correlator (ALV GmbH, Langen, Germany) and the digital input of the Virtex II FPGA device. The duration of the auto- and cross-correlation measurements was 5×20 s. ACF curves of the dye solutions were fitted to a model function assuming triplet state formation (Widengren et al., 1995) and one diffusion component:

$$G(\tau) = \frac{1}{N}\frac{1-T+Te^{\tau/\tau_{tr}}}{1-T}\left(1+\frac{\tau}{\tau_D}\right)^{-1}\left(1+\frac{\tau}{S^2\tau_D}\right)^{-1/2} \quad (10)$$

where $N$ is the average number of dye molecules in the detection volume, $\tau$ is the lag time, $T$ denotes the equilibrium molar fraction of fluorophores in the triplet state, $\tau_{tr}$ is the triplet lifetime, $\tau_D$ is the diffusion time, and $S$ is the ratio of the lateral and axial diameters of the ellipsoidal detection volume. ACF curves of the doubly labeled antibodies were fitted to a model with triplet state and two diffusion components in order to account for dyes attached to antibodies (slow component) and free, dissociated dyes (fast component):

$$G(\tau) = \frac{1}{N}\frac{1-T+Te^{\tau/\tau_{tr}}}{1-T}\left[r_1\left(1+\frac{\tau}{\tau_{D,1}}\right)^{-1}\left(1+\frac{\tau}{S^2\tau_{D,1}}\right)^{-1/2} + (1-r_1)\left(1+\frac{\tau}{\tau_{D,2}}\right)^{-1}\left(1+\frac{\tau}{S^2\tau_{D,2}}\right)^{-1/2}\right] \quad (11)$$

where $r_1$ and $1-r_1$ are the fractional amplitudes of the fast and slow species having diffusion times $\tau_{D,1}$ and $\tau_{D,2}$, respectively. The CCF curves of the antibodies were fitted with Eq. 10 with $T = 0$ because triplet formation of the two dyes are uncorrelated events. A fast diffusion component is not necessary because dissociated Alexa Fluor 488 and Cy5 molecules move separately, yielding no cross-correlation. Nonlinear fitting was performed by using the Quickfit 3.0 software (developed by Jan Krieger, http://www.dkfz.de/Macromol/quickfit).

The raw correlation functions calculated by the ALV correlator and the FPGA correlator overlap perfectly (FIG. 5) and yield practically identical fit parameters (TABLE V). These results demonstrate the correctness of calculations performed by the FPGA correlator and its applicability to calculate two ACF and two CCF curves in real-time.

## V. Conclusion

We presented new FPGA correlator designs to calculate auto- and cross-correlation functions using the multiple-τ scheme. The designs are based on sequentially scheduling the calculation of different time segments of the correlation functions, thus implementation of a single block with 8 correlation channels is sufficient to calculate the full correlation curve.

At the cost of increasing the minimal lag-time by a factor of 4, two ACFs and two CCFs are calculated in real-time from the photon streams of two APDs using the same number of correlation channels. Raw intensity traces with high time resolution are also stored by the design allowing post-acquisition data analysis (filtering of data, photon counting histogram analysis, etc.). The LabVIEW code can be used in many other types of FPGA cards produced by National Instruments. The bitfiles and LabVIEW source codes are available on request.

The concept of scheduling can be applied on any other FPGA card using other programming languages such as VHDL. The number of calculated correlation functions can easily be increased by implementing more CorrPEs. As shown in the accompanying paper[13], 1024 autocorrelation functions of data from a 32×32 APD array were calculated simultaneously. Further applications in FCS, including computation of auto- and cross-correlation functions from several discrete APDs or control of experiments with alternating laser excitation, should be easily implemented using our design.

**Acknowledgements**

This work was supported by grants from the Hungarian Scientific Research Fund OTKA K77600, MOLMEDREX (FP7-REGPOT-2008-1. #229920), the German-Hungarian program for the exchange of researchers by the German Academic Exchange Service and the Hungarian Scholarship Board (MÖB-47-1/2010) and TÁMOP-4.2.1/B-09/1/KONV-2010-0007 to GV implemented through the New Hungary Development Plan co-financed by the European Social Fund and the European Regional Development Fund. We thank Dr. László Tóth (Department of Informatics Systems and Networks, University of Debrecen) for initiating the collaboration between the authors and giving advice on the use of FPGAs, and Ferenc Nagy for suggesting a solution for calculating $\log_2 X_c$ described in the Appendix B.

## APPENDIX A: Proposition and proof related to Eq. 6

The order of execution of correlator blocks satisfying the requirements of the multiple-τ scheme, i.e., the (s+1)-th block can be executed after the s-th block has been executed twice, was given by the sequence defined by Eq. 6. We show that this sequence fulfills this requirement. The $s_c$ sequence is defined as:

$$s_c = \begin{cases} 0 & \text{if } c \text{ is odd} \\ s_{c/2} + 1 & \text{otherwise} \end{cases} \quad c \in \mathbb{N}\{0\} \quad \text{(A1)}$$

Let $b$ and $d$ be values of $c$ at which the (s+1)-th block is processed two consecutive times ($s_b = s_c = s+1$). Let $m$ and $p$ be indices at which the s-th block is processed two consecutive times ($s_m = s_p = s$).
If there are exactly two integers $p$ and $m$ such that $b < m < p < d$, then Eq.A1 fulfills the requirement of the order of executions of blocks.

Statement:
There are exactly two numbers: $p, m \in \mathbb{N}\setminus\{0\}$ such that $b < m < p < d$.

Proof:
$b$ can be written uniquely with s+1 and $n \in N : \{1,3,5,7,9...\}$ in the form:

$$b = 2^{s+1} \cdot n \quad \text{(A2)}$$

The next odd number after $n$ is $n+2$, thus the next index where the s+1-th block is processed:

$$d = 2^{s+1}(n+2) \quad \text{(A3)}$$

$m$ can also be written uniquely with $s \in \mathbb{N}$ and $h \in N$ such that

$$m = 2^s h \implies p = 2^s (h+2) \quad \text{(A4)}$$

$\exists f \in \mathbb{Z}$ such that $h = n + 2f$, thus

$$m = 2^s(n+2f) \implies p = 2^s(n+2(f+1)) \quad \text{(A5)}$$

If $f$ is such that $b < m < p < d$ then the s-th block was executed twice between two consecutive executions of the (s+1)-th block.

$$b < m \implies n < 2f \quad \text{(A6)}$$
$$p < d \implies 2f < n+2 \quad \text{(A7)}$$

From Eqs A1.6 and A1.7 it follows that

$$n < 2f < (n+2) \implies f = (n+1)/2 \quad \text{(A8)}$$

If the s-th block were executed a third time before $d$, then $\exists q = 2^s(h+2+2)$ such that

$$q < d \implies 2f < n \quad \text{(A9)}$$

which is in contradiction with Eq. A.6. Therefore the s-th block is executed only twice between two constitutive executions of the (s+1)-th block. (q.e.d.)

## APPENDIX B: Derivation of Eq. 7

From the previous derivation it can also be proven that if $s_c = k$ then $s_{c+2^{k+1}} = k$, i.e., the occurrence of the value $k$ in the sequence is periodic.

We have shown above that $s_c = s_{2^k \cdot n} = k$ for any c. Because $k$ is the exponent of 2, $k$ marks the position of the first "1" digit in the binary form of $c$.

Let us suppose that the position of the first "1" digit is $s \in \mathbb{N}$ in the binary form of (32-bit unsigned integer) $c$. The higher digits can be any $\alpha_i$:

$$c = (\alpha_{32}\alpha_{31}...\alpha_{s+1}1_s 00....0)_2 \tag{B1}$$

The following formula was used to find this position $s$ (Eq. 7 in the text):

$$Log_2\left(\left((c\ XOR\ (c-1))+1\right)/2\right) \tag{B2}$$

The result of A1.11 is equal to $s$ because:

$$c-1 = (\alpha_{32}\alpha_{31}...\alpha_{s+1}0_s 11....1)_2 \tag{B3}$$

$$c\ XOR\ (c-1) = (00...0_{s+1}1_s 11....1)_2 \tag{B4}$$

$$(c\ XOR\ (c-1))+1 = (00...1_{s+1}0_s 00....0)_2 \tag{B5}$$

$$X_c = \frac{1}{2}\left[(c\ XOR\ (c-1))+1\right] = (00...0_{s+1}1_s 00....0)_2 \tag{B6}$$

$$(s)_{10} = Log_2(X_c) \tag{B7}$$

$X_c$ defined in Eq. B6 is always a power of 2, therefore calculating $log_2 X_c$ is equivalent to finding the position of the only "1" digit in the binary form of $X_c$. The next procedure is used to find this position in $X_c$.

Step 1

$$m_4 = x\ AND\ (11111111111111110000000000000000)_2 \tag{B8}$$

$$m_3 = x\ AND\ (11111111000000001111111100000000)_2 \tag{B9}$$

$$m_2 = x\ AND\ (11110000111100001111000011110000)_2 \tag{B10}$$

$$m_1 = x\ AND\ (11001100110011001100110011001100)_2 \tag{B11}$$

$$m_0 = x\ AND\ (10101010101010101010101010101010)_2 \tag{B12}$$

where "$p\ AND\ q$" is a bitwise AND operation between $p$ and $q$.

Then the $m_i$ values are the binary representation of $s$:

$$s = (m_4 m_3 m_2 m_1 m_0)_2 = \sum_{i=0}^{4} m_i \cdot 2^i \tag{B13}$$

**APPENDIX C: Derivation related to Eq. 8**

Statement: the index at which the $s$-th block must be executed for first time is

$$c_s^{(threshold)} = 19 \cdot 2^s - 16 \tag{C1}$$

Proof:
Let $b$ and $d$ be the indices at which the $(s-1)$-th block is executed for the $(L+m)$-th and $(L+m+1)$-th time (for the definition of $L$ and $m$ see the text). Then

$$b < c_s^{(threshold)} < d \tag{C2}$$

We write $c_{s-1}^{(threshold)}$ in the following form with $n \in N:\{1,3,5,7,9...\}$:

$$c_{s-1}^{(threshold)} = 2^{s-1} \cdot n \tag{C3}$$

Then $b = 2^{s-1}(n + 2 \cdot (L+m-1))$ (C4)

and $d = 2^{s-1}(n + 2 \cdot (L+m))$ (C5)

We write $c_s^{(\text{threshold})}$ in the following form with $h \in N$:

$$c_s^{(\text{threshold})} = 2^s \cdot h. \tag{C6}$$

Then $\exists g \in \mathbb{Z}$ such that $c_s^{(\text{threshold})} = 2^s (n + 2 \cdot g)$. (C7)

It follows from Eqs. C2 and C7 that

$$d > c_s^{(\text{threshold})} \Rightarrow -n + 2 \cdot (L + m) > 4 \cdot g \tag{C8}$$

and

$$c_s^{(\text{threshold})} > b \Rightarrow 4 \cdot g > -n + 2 \cdot (L + m - 1). \tag{C9}$$

Eqs. C8 and C9 are satisfied if

$$g = (-n + 2 \cdot (L + m) - 1)/4. \tag{C10}$$

Thus, $c_s^{(\text{threshold})} = 2^s (-\frac{1}{2} + L + m + \frac{1}{2} n)$. (C11)

From Eqs. C3 and C11 we can write:

$$d_s = c_s^{(\text{threshold})} - c_{s-1}^{(\text{threshold})} = 2^{s-1}(2 \cdot (L + m) - 1) \tag{C12}$$

If we add the differences between the first executions of the different blocks, we get $c_s^{(\text{threshold})}$:

$$c_s^{(\text{threshold})} = c_0^{(\text{threshold})} + \sum_{i=1}^{s} d_i = c_0^{(\text{threshold})} + (2^s - 1) \cdot (2 \cdot (L + m) - 1) \tag{C13}$$

In our design $L=8$, $m=2$ and the 0-th block is executed first at $c_0^{(\text{threshold})} = 3$, thus:

$$c_s^{(\text{threshold})} = 19 \cdot 2^s - 16. \tag{C14}$$

**Tables**

**TABLE I.** Execution order of the CorrPE-1 design

| c | 1 | 2 | 3 | 4 | 5 | 6 | 7 | 8 | 9 | 10 | 11 | 12 |
|---|---|---|---|---|---|---|---|---|---|---|---|---|
| $s_c$ | 0 | 1 | 0 | 2 | 0 | 1 | 0 | 3 | 0 | 1 | 0 | 2 |
| Executed block | -[1] | - | 0 | - | 0 | - | 0 | - | 0 | - | 0 | - |
| c | 13 | 14 | 15 | 16 | 17 | 18 | 19 | 20 | 21 | 22 | 23 | 24 |
| $s_c$ | 0 | 1 | 0 | 4 | 0 | 1 | 0 | 2 | 0 | 1 | 0 | 3 |
| Executed block | 0 | - | 0 | - | 0 | - | 0 | - | 0 | 1 | 0 | - |
| c | 25 | 26 | 27 | 28 | 29 | 30 | 31 | 32 | 33 | 34 | 35 | 36 |
| $s_c$ | 0 | 1 | 0 | 2 | 0 | 1 | 0 | 5 | 0 | 1 | 0 | 2 |
| Executed block | 0 | 1 | 0 | - | 0 | 1 | 0 | - | 0 | - | 0 | - |

[1] Hyphens (-) denote empty executions

**TABLE II.** Order of execution of blocks with different values of $c_{corr}$ and $c$ in the CorrPE-2 design

| $c_{corr}$ | 0 | 1 | 2 | 3 | 0 | 1 | 2 | 3 | 0 | 1 | 2 | 3 |
|---|---|---|---|---|---|---|---|---|---|---|---|---|
| selected correlation function[a] | $g^x$ | $g^y$ | $g^{xy}$ | $g^{yx}$ | $g^x$ | $g^y$ | $g^{xy}$ | $g^{yx}$ | $g^x$ | $g^y$ | $g^{xy}$ | $g^{yx}$ |
| C | 30 | 30 | 30 | 30 | 31 | 31 | 31 | 31 | 32 | 32 | 32 | 32 |
| $s_c$ | 1 | 1 | 1 | 1 | 0 | 0 | 0 | 0 | 5 | 5 | 5 | 5 |
| Calculated elements[b] | $G_{1,i}^{(x,x)}$ | $G_{1,i}^{(y,y)}$ | $G_{1,i}^{(x,y)}$ | $G_{1,i}^{(y,x)}$ | $G_{0,i}^{(x,x)}$ | $G_{0,i}^{(y,y)}$ | $G_{0,i}^{(x,y)}$ | $G_{0,i}^{(y,x)}$ | - | - | - | - |

[a] $g^x$, $g^y$, $g^{xy}$, $g^{yx}$ denote the selected correlation function, which is calculated in the given cycle iteration. $g^x$ and $g^y$ are ACFs from detector $x$ and $y$. $g^{xy}$ and $g^{yx}$ are CCFs between the signals of detectors $x$ and $y$.

[b] Hyphens (-) denote empty executions.

**TABLE III.** Memory usage of the correlator elements

| Name of memory block | Representation | # of memory blocks | # of RAMB16-S36_S36[1] units | Memory usage[2] CorrPE-1/-2X | |
|---|---|---|---|---|---|
| Read-out block | 64-bit | 9 | 9x2 | 1.8 | 6.62 |
| $G^{(x,y)}_{s,i}$ | 64-bit | 8 | 8x2 | 1.6 | 5.88 |
| $I^{(d,y)}_{s,i}$ | 32-bit | 10 | 10 | 1 | 1.84 |
| $I^{(u,x)}_s, I^{(u,x)*}_s$ | 32-bit | 2 | 2 | 0.2 | 0.37 |
| $M^{(x)}_s, T_s$ | 32-bit | 2 | 2 | 0.2 | 0.28 |
| Total: | N.A. | 31 | 48 | 5.1 | 15 |

[1]RAMB16_S36_S36 is a Virtex-II dual-port RAM. The 64-bit representation is stored physically in two 32-bit blocks.
[2]Kbyte

**TABLE IV.** Device utilization of the compiled design of CorrPE-2X

| Name of element | Used | Available | Percent |
|---|---|---|---|
| Number of Slices | 5318 | 14336 | 37% |
| Number of BRAMs | 60 | 96 | 62% |
| Number of MULT18X18s | 45 | 96 | 46% |

**TABLE V.** Fit parameters of the auto- and cross-correlation functions

| | | Correlator | $T^1$ | $\tau_{tr}$ [µs] | $N$ | $r_1$ | $\tau_{D1}$ [µs] | $\tau_{D2}$ [µs] | $S$ |
|---|---|---|---|---|---|---|---|---|---|
| A488 | $g^x$ | CorrPE-1 | 0.26±0.02 | 1.6±0.2 | 15.5±0.1 | - | 26.6±0.4 | - | 7.1±0.5 |
| | $g^x$ | ALV | 0.28±0.04 | 1.2±0.2 | 15.0±0.1 | - | 25.4±0.5 | - | 7.7±0.9 |
| Cy5 | $g^y$ | CorrPE-1 | 0.49±0.01 | 1.8±0.1 | 16.4±0.1 | - | 44.8±0.5 | - | 6.8±0.4 |
| | $g^y$ | ALV | 0.48±0.01 | 1.9±0.1 | 16.2±0.1 | - | 43.7±0.6 | - | 8.0±0.9 |
| A488-Cy5-W6/32 | $g^x$ | CorrPE-2X | 0.15±0.01 | 2.0±0.2 | 16.5±0.1 | 0.33±0.0 | 27.0±2.4 | 256.2±6.4 | 8.1±1.2 |
| | $g^x$ | ALV | 0.14±0.01 | 2.3±0.3 | 16.4±0.1 | 0.34±0.0 | 27.2±2.6 | 256.3±4.9 | 9.0±1.9 |
| | $g^y$ | CorrPE-2X | 0.12±0.02 | 1.9±0.5 | 25.3±0.6 | 0.17±0.0 | 29.2±5.5 | 405.4±13 | 6.0±1.6 |
| | $g^y$ | ALV | 0.12±0.03 | 2.9±0.8 | 26.4±0.7 | 0.14±0.0 | 27.3±4.9 | 401.0±31 | 6.0±4.7 |
| | $g^{xy}$ | CorrPE-2X | - | - | 67.4±16 | - | - | 312.3±3.0 | 8.8±3.3 |
| | $g^{xy}$ | ALV | - | - | 65.8±1.6 | - | - | 328.8±9.1 | 8.6±4.1 |
| | $g^{yx}$ | CorrPE-2X | - | - | 68.1±3.4 | - | - | 358.8±3.9 | 8.4±4.7 |
| | $g^{yx}$ | ALV | - | - | 67.7±0.4 | - | - | 340.4±3.2 | 9.1±3.2 |

[1]Values are given as means ± s.d.

**Figures**

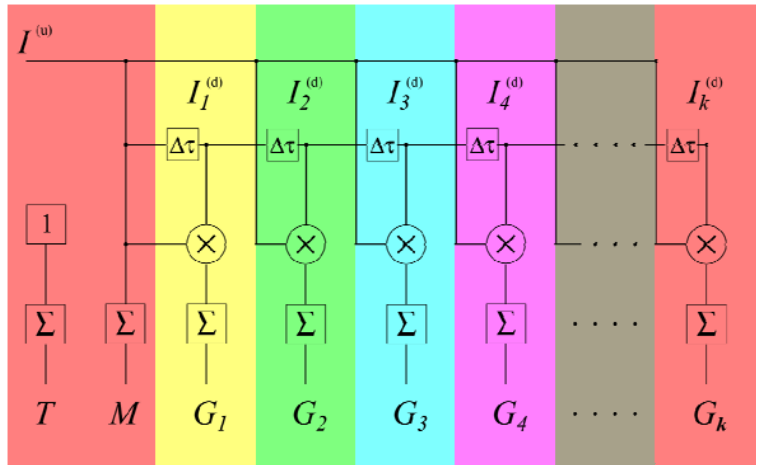

**FIG. 1** Data flow in a linear correlator
$I^{(u)}$ is the undelayed intensity, $I_i^{(d)}$ values are delayed intensities stored in a shift register. The $G_i$ values are calculated by correlation channels (see text). $M$ is the monitor channel, which stores the sum of the undelayed intensities. T counts the number of $\Delta\tau$ intervals.

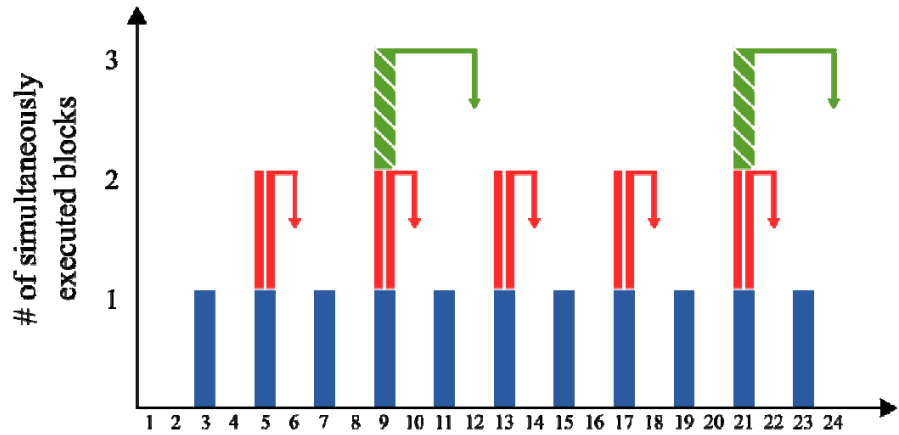

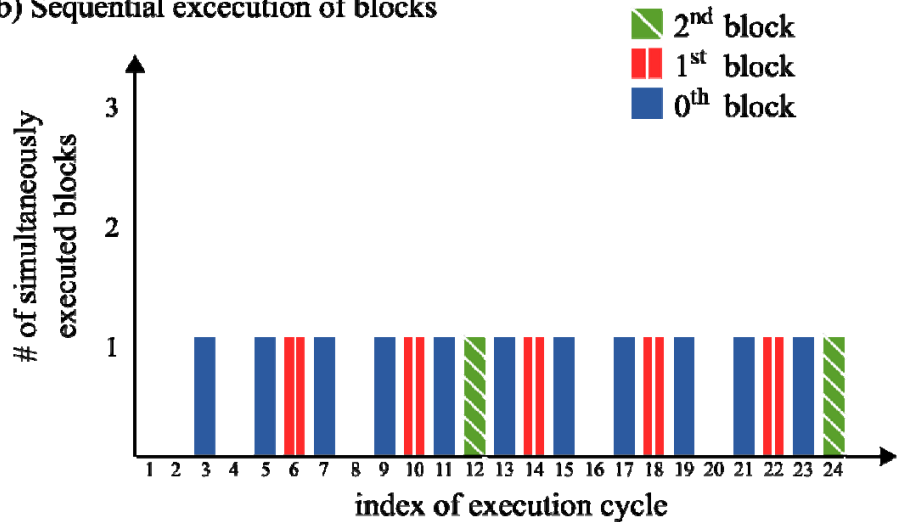

**FIG. 2** Order of execution of linear correlator blocks
The parallel execution of linear correlator blocks is shown in (a). The frequency of execution of blocks with longer lag-times decreases by powers of 2. In our design the linear correlator blocks are executed sequentially. Succeeding blocks are executed in the unused time slots between two executions of the $0^{th}$ block (Arrows in Panel a)). Panel b) shows the sequentially scheduled execution of linear correlator blocks. The index of the executed block is given by equations 7 and 8.

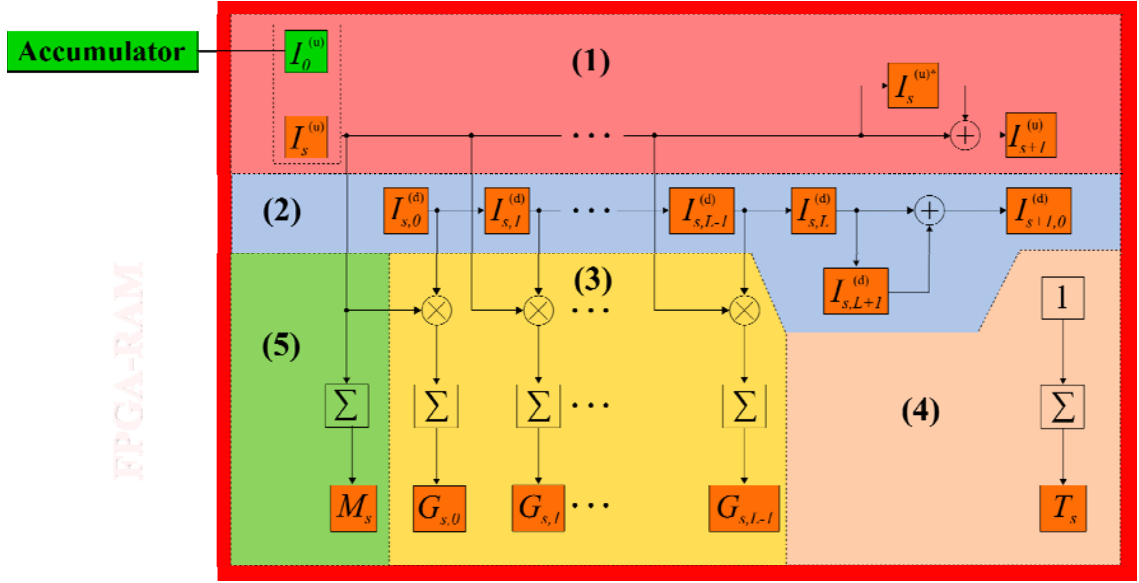

**FIG. 3** Logic circuit of the Correlator Unit
The operation of the Correlator Unit is described in the text; the numbered parts of the scheme refer to the points of the description. Parameters in orange boxes represent FPGA-RAM access.

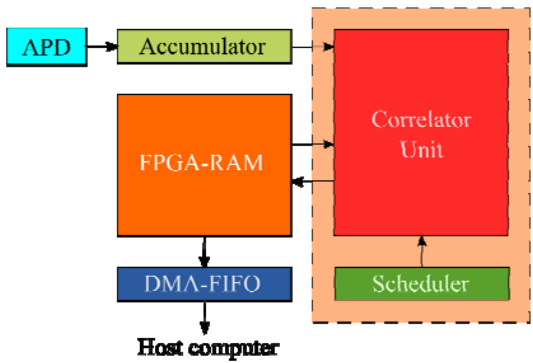
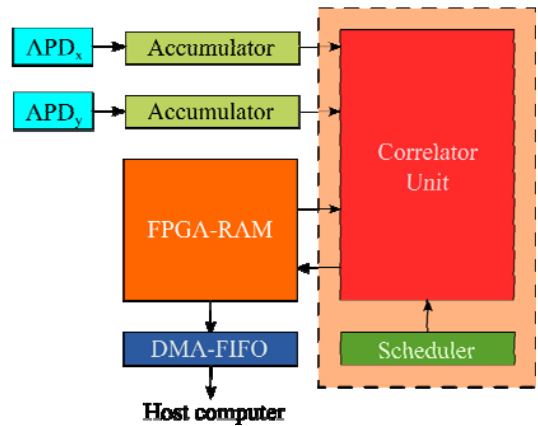

**FIG. 4** Complete designs for 1 or 2 input signals
CorrPE-1 has one input signal, whereas CorrPE-2X collects input signals from two different APDs. Designs consist of the following elements: accumulator(s) summing up the incoming photon counts from the APD(s) between two sampling periods of the 0-th block; the FPGA-RAM storing all parameters of all blocks and the data for read-out; CorrPE-1 or CorrPE-2X made up from the correlator unit and the scheduler. The DMA-FIFO receives data from the read-out part of the FPGA-RAM, and sends them to the host computer.

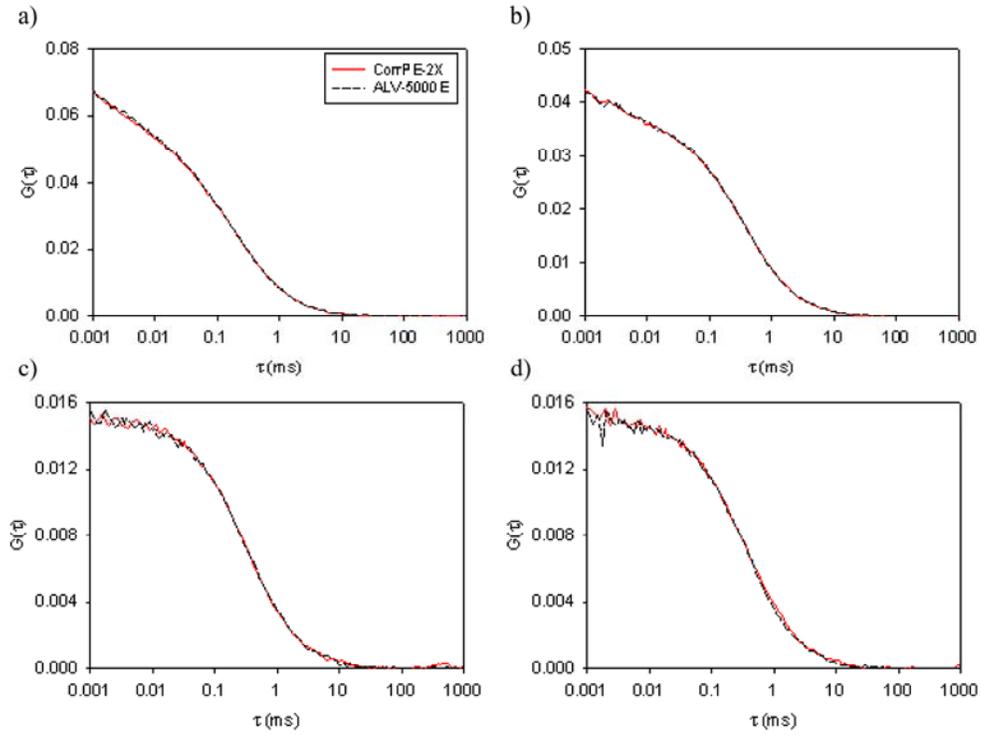

**FIG. 5** Validation of the FPGA correlator
Auto- and crosscorrelation curves of W6/32 antibodies double-labeled with Alexa Fluor 488 and Cy5 were calculated by CorrPE-2X on the FPGA (red) or the ALV-5000E hardware correlator (black). Panel a) and b) show autocorrelation curves of the green signal ($APD_x$) and the red signal ($APD_y$), respectively. Panels c) and d) are the crosscorrelations of the two signals.